# Social Media Marketing (SMM) – A Strategic Tool for Developing Business for Tourism Companies

Dr. Nalini Palaniswamy*

*Assistant Professor,

Centre for Marketing,

KCT Business School,

Kumaraguru College of Technology,

Coimbatore, India.

## Abstract

Social media marketing is an emerging marketing technique worldwide. This research concentrates on how effectively social media can be used to promote a product in tourism industry. The efficient use of social media develops a tourism company in terms of sales, branding, reach and relationship management. The study aims to find the best social media platform to promote and develop a tourism company and the customer opinion towards planning a trip through online. It also concentrates on customer response for online offers and discounts in those social media platforms. The study attempts to understand and create suitable model for social media marketing for tourism companies with a sample size of 400. The sampling technique used in this study is purposive sampling method. The purposive sample can also be called as judgemental sample. Normally the sample will be selected based on the knowledge possessed by the respondents on a particular phenomenon. Here, the study is been conducted among the people who use social media. The sampling technique helped the researcher to identify the target sample i.e., the social media users.

SMM is an internet based tool used to inform, persuade, attract, and educate the present and potential customers through online. The millennials by enlarge is understood as tech savys  they spend most of their time in internet based apps and social communities. They use Social media to interact, share ideas, emotions, experience and so on. The social media also reinvented the word of mouth communication into word of internet and also the viral communication is very fast when compared to other traditional Medias. So as a marketers in today's scenario we have to understand how this medium can be well utilised for business. This study attempts to understand the how SMM can be effectively used for the tourism companies. The study can be tested widely across various tourism companies and it can validated  and the study results can be generalised.

___________________________________________________________________________

## Introduction

Social media marketing is an emerging marketing technique worldwide. The efficient use of social media develops a tourism company in terms of sales, branding, reach and relationship management. The research concentrates on how effectively social media can be used to promote a product in tourism companies. The research paper portrays the best social media platform to promote and develop a tourism company and the customer opinion towards planning a trip through online. It also concentrates on customer response for online offers and discounts in those social media platforms. The social media plays an important role in spreading the information through social communities and SMM is a way of marketing products and services via online communities, Blogs, social networks and SEOs. Marketing a company's offerings and products through social media is widely called as social Media marketing(SMM).

## Social Media Marketing in Tourism Companies

Today's consumers are influenced very much by the social media. When the travellers plan a vacation they start surfing information via online instead of getting personal contacts with tourism companies. This is a great advantage for the tourism companies to invest more on developing social media sites and minimise the contact personnel for the companies. In comparison with the collected data it was found that the consumers are so much engaged in various social media sites like Facebook, Youtube, E-mails, Twitter, and Linkedin to gather, and share the information related to their travel destinations. The SMM in tourism companies is used in two ways, one to surf information about the tourism places and its features, two to share photos, videos and experiences to the social communities.

There are various benefits enjoyed by the travel companies because of this growing social media like many of the disadvantages of traditional medium is replaced by social media. The travel companies can target the different demographics individually. In addition to it, the SMM helps to study and understand individual consumer interest, behaviour pertaining to various psychological factors. Another great advantage of SMM is they ROI. The retur on investment is highly measureable and investment on product promotion can be controlled. SMM is not new anymore but the usage and tapping of potential market is the underlying challenge for the tourism companies.

India is been ranked third among fastest growing tourism sectors. India is also predicted to be the large medical tourism sector with estimated growth of 30 percent annually and also estimated to reach a good picture of 95 billion in 2017. Overall growth of the industry is predicated as 7.9% from 2014-2025[*]

___________

[*] Indian tourism industry analysis report 2014

## Review of Literature

The social media utilisation has remarkable improvements. The consumers are utilising social media for content sharing, blogs, social networking and wikis. The activities done in this different mediums are not same. There is a considerable difference between this mediums (Janh. Kietzmann 2011), Social media plays a vital role for travel related searches. The study results of Zheng Xiang (2010) showed how much the social medium constitute for search engines sites. The study provided a clear evidence for various challenges faced by tourism travellers and how social media helped them in their travel. The social media sites has enabled business to build closer relationship with their customers, as well expand the total customer base for the companies. Kevin J Trainoor etal. (2014), shows how usage of social media marketing can be customer centric and how usage of social media marketing can contribute for capacity building at firm level. The firm level objectives can be attained by applying a customer centric approach through various social media sites.

## Objectives of the Study

The study was carried out to understand the following objectives

- To examine the profile of social media users with reference to tourism companies

- To know the consumer preference in usage and selection of various social media sites

- To understand the relation between the various marketing strategies using social media sites and its impact on consumer decision making

## Methodology

The period of study was starting from January 2015 to June 2015. The study attempts to understand and create suitable marketing strategies via social media marketing for Tourism Companies with a sample size of 400. The sampling technique used in this study is purposive sampling method. The purposive sample can also be called as judgemental sample. Normally the sample will be selected based on the knowledge possessed by the respondents on a particular phenomenon. Here, the study is been conducted among the people who use social media. The sampling technique helped the researcher to identify the target sample the social media users.The sampling technique helped the researcher to identify the target sample i.e the social media users. The study data was validated and with the reliability of more than 7 it was tested using various analysis to support its hypothesis. The test used for the study includes ANOVA, Garett ranking and Correlation

## Profile of the Respondents

The table 1 portrays the profile of the respondents for the study. The majority of the respondents belongs to the age group of 20-30 who are also identified as internet/tech savvy because internet and social media has become part of their life style. Most of the social media and online users are males. The Professional degree holders seems to be highest in preferring social Medias and the private employees are preferring the online media than the other segments. The respondents using the social media and online sources earn more than 5 lakhs per annum.

## Findings and Discussions

The growing dual income families and raising income lead its way for development of tourism and travels. The lifestyle modification and growth of internet usage and social networking plays an important role in today's growth of this tertiary sector.

The study results give very interesting picture about the preference of the social media users in selecting their tours and travels. 43.3% of respondents are willing to go for adventure kind of tours every 6 months. Almost 55.6% of respondents has already gone for overseas trip using the social media and online apps. Out of the total respondents about 60% of respondents have already planned trip through online and they feel it to be the comfortable mode to plan and execute their travel in orderly fashion.

### Table: 1 Profile of the Respondents

| Age | Frequency | Percentage |
|---|---|---|
| **20-30** | **198** | **49.5** |
| 30-40 | 139 | 34.5 |
| 40-50 | 54 | 13.5 |
| 50-60 | 9 | 2.5 |
| **Gender** | **Frequency** | **Percentage** |
| Male | 232 | 58 |
| Female | 168 | 42 |
| **Education** | **Frequency** | **Percentage** |
| School level | 8 | 2.0 |
| UG | 56 | 14.1 |
| PG | 153 | 38.2 |
| Professional | 183 | 45.7 |
| Occupation | Frequency | Percentage |
| Govt employee | 66 | 16.5 |
| Private employee | 102 | 25.5 |
| Students | 35 | 8.75 |
| Self employed | 45 | 11.25 |
| Business | 92 | 23.0 |
| Professional | 60 | 15 |
| Income | Frequency | Percentage |
| Below 5 lac | 125 | 31.2 |
| 5 lacs - 10 lacs | 107 | 26.7 |
| 10lacs - 15lacs | 97 | 24.2 |
| Above 15lacs | 71 | 17.9 |

The respondents using the social media view different things according to their interests and usage. Out of 400 respondents 25% of respondents have seen offers and discounts in various pages which gives information of tours and travels, 22.5% of respondents view posts and videos, 20% of respondents have view banner ads and 16.3% of respondents view events and updates on various conferences, business meets, sports etc., It is understood that people see different things when they look for travel. Since the online forum is very comfortable medium for comparing the service providers the results show that 55.6% of respondents compare two companies before planning a trip and 42.5% of respondents highly prefer online offers & discounts.

Promotions , offers, travel packages are vital part of this service sectors, the results show that out of the total respondents studied 37% of respondents are impressed by Gift voucher,27.4 % of respondents are impressed by promotional offers, 21.8% of respondents are impressed by add ons. Hence the majority prefers Gift voucher. The respondents prefer promotion messages like photos & videos are highly influential in attracting the new and exciting customers. There was a considerable difference between the customer's preference towards planning a trip through online and offline. The tourism companies should concentrate on farming different marketing strategies for online and offline companies.

## The Ranking of Preference of Social Media Platforms as per Accessibility

In general the social media is viewed as powerful medium in the current market space. In spite there is internal preference when SMM is listed.

### Table No-2 -Table Showing the Ranks of the Respondent's Present Accessibility in various Social Media

| S.no | Social Media Platform | Total score | Mean score | Rank |
|---|---|---|---|---|
| 1 | E-mail | 9882 | 47.36 | 3 |
| 2 | Facebook | 10345 | 55.24 | 1 |
| 3 | Youtube | 9900 | 50.02 | 2 |
| 4 | Linked in | 8896 | 39.45 | 5 |
| 5 | Twitter | 9108 | 43.67 | 4 |

The social media in general classified in to various platforms. The study results has ranked the respondent's accessibility of various platforms the results shows that E-mail, Facebook and YouTube tops. The tourism companies should concentrate on the above items to promote their services at first level.

## Preference of Social Media Platforms as per the respondent's Interest

It was found that there is a considerable difference between the accessibility and usage of Social media the following ranks clearly picturizes the individual interest on using the social media

### Table No-3-Table showing the Interest of respondents in using the Social Media

| S.no | Social media platform | Total score | Mean score | Rank |
|---|---|---|---|---|
| 1 | E-mail | 9882 | 47.36 | 3 |
| 2 | Facebook | 10345 | 55.24 | 1 |
| 3 | Youtube | 9900 | 50.02 | 2 |
| 4 | Linked in | 8896 | 39.45 | 5 |
| 5 | Twitter | 9108 | 43.67 | 4 |

The table 3 shows the interest of the respondents of various social media platforms. We can see that the rank of preference of social media sites as per interest. Facebook ranks first, YouTube ranks second, E-mail ranks third, . It is very clear people like to see updates via Facebook than other Medias.

**Preference of Social Media Platforms for Networking**

### Table 4- The Media usage as per the usage of Networking

| S.no | Social media platform | Total score | Mean score | Rank |
|---|---|---|---|---|
| 1 | E-mail | 9882 | 47.36 | 3 |
| 2 | Facebook | 10345 | 55.24 | 1 |
| 3 | Youtube | 9900 | 50.02 | 2 |
| 4 | Linked in | 8896 | 39.45 | 5 |
| 5 | Twitter | 9108 | 43.67 | 4 |

From the above table, we can see that the rank of usage of social media sites as per Better networking. Facebook ranks first, Youtube ranks second, E-mail ranks third, Twitter ranks fourth and linked in ranks fifth. Respondents prefer to network in order to understand about their travel and places they like to explore. This mediums can be prioritize when the tourism companies target on educating its potential customers.

**Social Media Platforms in Information Dissemination**

The respondent's views on information dissemination in various social media websites are as follows

### Table 5 -Table Showing the Social Media Platforms in Information Dissemination

| S.no | Social media platform | Total score | Mean score | Rank |
|---|---|---|---|---|
| 1 | E-mail | 9882 | 47.36 | 3 |
| 2 | Facebook | 10345 | 55.24 | 1 |
| 3 | Youtube | 9900 | 50.02 | 2 |
| 4 | Linked in | 8896 | 39.45 | 5 |
| 5 | Twitter | 9108 | 43.67 | 4 |

Table 5, picturize the information dissemination is very good in Facebook followed by you tube and e-mails. Twitter and Linked in ranks the last. The tourism companies can design the message in the top ranked media instead concentrating on the least ranked items.

**Correlation**

The correlation test was used to measure the strength and direction of association between two variables namely customer who sees social media advertisements and customer's preference towards planning a trip through online. The pearson coefficient r, is 0.985 which is greater than 0.05. Hence there is a strong relationship between preference of respondents towards planning a trip through online and the advertisements that they see in social media. The pearson r, value is positive which shows that the one variable increases in value, the other value will also increase. Hence the customer who sees more social media advertisements will increase the customer's preference towards planning a trip through online.

Another correlation test was used to measure the strength and direction of association between two variables i.e., preference of online offers and discounts and respondents who is interested in planning a trip when they find a good offer From the above table, the pearson coefficient r, is 0.701 which is greater than 0.05. Hence there is a strong relationship between preference of online offers and discounts and respondents who is interested in planning a trip when they find a good offer. The pearson r, value is positive which shows that the one variable increases in value, the other value will also increase. Hence the customer who prefer offers & discounts will plan a trip when they find a good offer.

## Conclusion

SMM is not a new concept but the tourism companies should take a new way to tap its existing potential.  SMM is a very distinct way of promoting a company's information to its potential customers so utmost important is needed for promoting the information via this viral medium. The tourism companies started working on various tourism products perhaps there should an equal concentration on growing medium is important to tap their business. As per the statistics of social media users India is at top 5 and it is another attracting factors. Free internets and lower internet costs will penetrate the segments still wider in future years. Now it is the time for the companies to change, update and adopt to the new changing technologies.

## Scope of further Study

The study concentrates on one of the module of digital marketing i.e., social media marketing. In future the entire digital marketing can be studied. Social media marketing will help to study the Strategic planning in digital marketing arena.